\documentclass[aps,prl,showpacs,twocolumn,superscriptaddress]{revtex4}
\usepackage{amssymb}

\usepackage{amsmath}
\usepackage{bm}
\usepackage{graphicx}

\begin{document}

\title{Crossover from Kramers to phase-diffusion switching in hysteretic DC-SQUIDs}
\author{J. M\"annik}
\affiliation{Department of Physics and Astronomy, Stony Brook University, Stony Brook, NY
11794}
\author{S. Li}
\affiliation{Department of Physics and Astronomy, University of Kansas, Lawrence, KS 66045}
\author{W. Qiu}
\affiliation{Department of Physics and Astronomy, University of Kansas, Lawrence, KS 66045}
\author{W. Chen}
\affiliation{Department of Physics and Astronomy, Stony Brook University, Stony Brook, NY
11794}
\author{V. Patel}
\affiliation{Department of Physics and Astronomy, Stony Brook University, Stony Brook, NY
11794}
\author{S. Han}
\affiliation{Department of Physics and Astronomy, University of Kansas, Lawrence, KS 66045}
\author{J. E. Lukens}
\affiliation{Department of Physics and Astronomy, Stony Brook University, Stony Brook, NY
11794}
\date{\today}

\begin{abstract}
We have measured and propose a model for switching rates in hysteretic
DC-SQUID in the regime where phase diffusion processes start to occur. We
show that the switching rates in this regime are smaller than the rates
given by Kramers' formula due to retrapping of Josephson phase. The
retrapping process, which is affected by the frequency dependent impedance
of the environment of the DC-SQUID, leads to a peaked second moment of the
switching distribution as a function of temperature. The temperature where
the peaks occur are proportional to the critical current of the DC- SQUID.
\end{abstract}

\maketitle

The phenomenon of phase-diffusion in Josephson junctions has been
extensively studied over the past 20 years \cite{Martinis,Iansiti, Vion,
Koval}. These studies, in general, focused on deep sub-micron junctions with
very low critical currents, $I_{c},$ so that the Josephson coupling energy $%
E_{J}$ was of the order $k_{B}T$ . Extensive diffusion of the phase
typically occurred before the junction switched to the so-called running
state with voltage near the superconducting gap. With the advent of research
on flux qubits for quantum computing using SQUID magnetometers with much
larger, unshunted junctions for readouts ($I_{c}\gtrsim 1\mu $A$)$, this
phase-diffusion process has reappeared, rather dramatically, in a different
regime. In these larger junctions, which are the focus of this work, the
crossover from simple Kramers activation to phase-diffusion before switching
tends to occur at higher temperatures and involves only a small number of
phase slip events before the switching process. Nevertheless the effects of
this crossover can be quite pronounced, appearing, perhaps most
dramatically, as a significant narrowing of the switching distribution and
thus, in a sense, a greatly increased sensitivity of the magnetometer,
albeit, at the expense of greater back action on the qubit.

The standard picture for the dynamics of a Josephson junction, shown in Fig.
1a, is that of a particle with a mass proportional to the junction
capacitance moving in a potential $U(\varphi )=-E_{J}\cos (\varphi
)-I_{b}\varphi ,$ where $\varphi $ is the phase and $I_{b}$ is the bias
current of the junction. For sufficiently low damping around the plasma
frequency ($\omega _{p}$) of the Josephson junction, the energy gain as the
particle moves from one barrier to the next will exceed the loss due to
damping. So, switching from the supercurrent state to running state is
triggered by a single event, the phase particle escaping from a potential
well, and no phase diffusion occurs. This situation is usually realized in
large area Josephson junctions. On the other hand, if the damping around the
junction plasma frequency is sufficiently high then at low bias the phase
particle will always retrap in a local minima after escape and a finite
resistance phase-diffusion branch exists on the I-V curve of the junction.
This case is usually realized in ultra-small Josephson junctions and can be
observed down to lowest temperatures attainable with a dilution refrigerator %
\cite{Vion}. For low damping, the switching rate equals to the thermally
activated (TA) escape rate from the potential minima which is given by well
known Kramers' formula \cite{Kramers}%
\begin{equation}
\Gamma _{T}=\frac{\Omega }{2\pi }a_{t}\exp \left( {-\frac{\Delta U}{k_{B}T}}%
\right) 
\end{equation}%
where $\Omega $ is the effective attempt frequency $(\Omega \approx \omega
_{p})$, $\Delta U$ is the height of potential barrier from a local minimum
to the next maximum (see also Fig. 1a) and $a_{t}$ is a damping dependent
factor ($0<a_{t}<1$) \cite{Li}. For the case of high damping, where phase
diffusion is always present at low bias, a similar expression can be given
for the switching rate although the meaning of the potential barrier and
attempt frequency are very different \cite{Vion}. For intermediate damping a
third regime occurs that bridges the previous two. At lower temperatures a
single escape can trigger the switching of the junction from the metastable
minima but at higher temperatures, where the mean escape current $\overline{%
I_{b}}$ is reduced, the energy gain is less and phase diffusion occurs. In
this case the switching rate is not only determined by the escape rate but
also by the probability of retrapping of the phase particle after it has
escaped.

The total transition rate from metastable minima to running state can be
written as a sum of rates through different escape and retrapping sequences
as $\Gamma =\sum_{n=0}^{\infty }p^{(n)}\Gamma ^{(n)}$. Here $\Gamma ^{(n)}$
stands for the switching rate of a process where the particle retraps $n$
times before the runaway starts and $p^{(n)}$ for the probability of such a
process. $p^{(n)}$ can be expressed in terms of the retrapping probability ($%
P_{rt}$), which we define as the probability that after the ascent of the
phase particle to a local maximum it will be retrapped in one of the
subsequent minima $p^{(n)}=(1-P_{rt})\cdot (P_{rt})^{n}$ . For the case
relevant to our data, where $\Gamma _{T}\ll \omega _{p},$ this gives

\begin{equation}
\Gamma =\Gamma _{T}(1-P_{rt})\sum_{n=0}^{\infty }\frac{(P_{rt})^{n}}{n+1}%
=\Gamma _{T}\left( 1-P_{rt}\right) \frac{\ln (1-P_{rt})^{-1}}{P_{rt}}
\end{equation}%
In Eq. 2, the switching rate factorizes into two parts, one of which is the
Kramers rate (1) and the other a function of the retrapping probability.
Since the latter is always less than or equal to one, the actual switching
rate is smaller than or equal to the rate predicted by Kramers equation (1).
At zero temperature, consideration of the balance between energy gain from
the bias source and energy losses due to the damping leads to the conclusion
that $P_{rt}(I_{b})$ will abruptly switch from 1 to 0 at some bias current $%
I_{r}-$the retrapping current. For ohmic damping and $Q\gg 1$ ($Q=\omega
_{p}CR),$ $I_{r\Omega }=4I_{c}/\pi Q$ \cite{BenJacob,Tinkham}. 
\begin{figure}[t]
\centerline{\includegraphics[width=70mm]{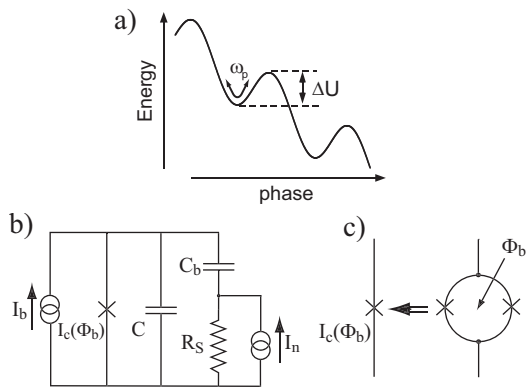}}
\caption{a) Schematics of washboard potential of DC-SQUID. Angular frequency
of small oscillations of the fictitious phase particle is $\protect\omega %
_{p}$. To escape phase particle has to overcome potential barrier $\Delta U$%
. b) Schematics of circuit used in calculation of retrapping probability. $%
R_{s}$, $C_{b}$ model high frequency damping seen by a Josephson junction
with critical current $I_{c}$. $I_{n}$ represents noise and $I_{b}$ bias
currents, respectively. $C$ marks capacitance of the junction. c) In
experiments single junction is replaced by a DC-SQUID which effective
critical current $I_{c}(\Phi _{b})$ can be modulated by external magnetic
field. }
\end{figure}

To find $P_{rt}$ for finite temperatures and frequency dependent damping, we
rely on Monte Carlo simulations. We calculate $P_{rt}$ for the RCSJ model
with a series RC shunt added to account for the stronger high frequency
damping (Fig. 1.b), e.g. due to the leads. Extensive discussion of this
model along with equations of motion for the phase can be found in ref. \cite%
{Kautz}. The damping at $\omega _{p}$ is assumed to be much stronger than
that at low frequency as required for the coexistence of phase-diffusion and
hysteresis. However, the exact value of the low frequency damping is not
important here, so we omit the quasiparticle resistance and bias resistors
which are much larger than the typical transmission line impedance ($\sim
100\,\Omega )$ seen by the junction near $\omega _{p}.$ To obtain $P_{rt},$
the particle is initiated repeatedly at the top of a barrier and its
subsequent motion in the presence of Nyquist noise is monitored until it is
retrapped in one of the metastable minima or switches to the running state.

Figure 2 shows this calculated $P_{rt}$ as a function of bias current for a
range of temperatures that is important for our data. At zero temperature, $%
P_{rt}$ switches abruptly, as expected from energy balance. At higher
temperatures, however, the retrapping probability curve broadens and shifts
to higher bias currents with $I_{b}^{50\%}\propto T^{1/2}$ where $%
I_{b}^{50\%}$ is defined as $P_{rt}(I_{b}^{50\%})=0.5$. Even though the
overall shift is toward higher bias as $T$ increases, $P_{rt}(I_{b})$\ also
decreases slightly from unity for $I_{b}<I_{r}$ where $I_{r}$ is the actual
retrapping current for the circuit model in Fig. 1a, i.e. $%
I_{r}=I_{r}(R_{s},C_{b})$ calculated at $T=0$. For $I_{b}\lesssim I_{r}$,
the simulations show $1-P_{rt}$ decreases essentially exponentially as $I_{b}
$ decreases. For this to occur, the frequency dependence of the damping is
crucial. If thermal fluctuations give enough initial velocity to the
particle, the linear component of $\frac{\partial \varphi }{\partial t}$
will increase, lowering the velocity dependent damping and thereby
permitting switching even though $I_{b}<I_{r}$. Due to high probability of
retrapping the switching rate $\Gamma (I_{b})$ deviates substantially from
Kramers' rate $\Gamma _{T}(I_{b})$ in this range of bias currents. 

\begin{figure}[tbp]
\centerline{\includegraphics[width=82mm]{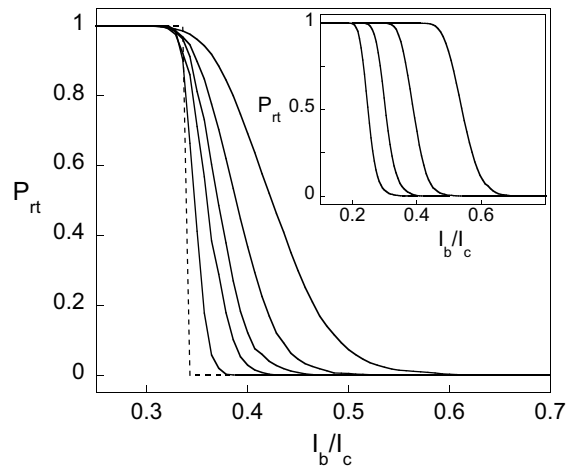}}
\caption{Retrapping probability as a function of normalized bias current for
circuit shown in Fig. 1. Parameters used in calculations are $I_{c}=2.90$ $%
\protect\mu $A, $C=260$ fF, $R_{s}=75$ $\Omega $, $C_{b}=15$ pF
corresponding to sample $B$. Curves from left to right correspond to
temperatures 0 (dashed line), 0.1, 0.5, 1.0, 2.0, 5.0 K. Inset shows
retrapping probability vs. normalized bias current for $R_{s}=125$, 100, 75,
50 $\Omega $ (from left to right) at $T=2.0$ K. $I_{c}$, $C$, $C_{b}$ have
been chosen the same as in the main Figure.}
\end{figure}

The effects of the choices of the circuit parameters $C_{b}$ and $R_{s}$
have been investigated using the simulations. The inset in Fig. 2 shows $%
P_{rt}(I_{b})$ for a range of $R_{s}$. As can be seen, $P_{rt}(I_{b})$\
shifts to lower bias currents with $I_{b}^{50\%}\propto 1/R_{s}$ as
predicted by $I_{r}\thickapprox 4I_{c}/\pi Q_{p}$ where for our non-Ohmic
circuit model $Q_{p}\equiv \omega _{p}CR_{s}$. We have chosen values of $%
C_{b}$ to be sufficiently large (15-20 pF) that the damping is nearly
frequency independent for frequencies \ near $\omega _{p}$. For these and
larger values of $C_{b},$ $P_{rt}(I_{b})$ is essentially independent of $%
C_{b}$ for $I_{b}>I_{r}.$ Only in the region $I_{b}<I_{r}$, does its exact
value become important. For determining the onset of phase-diffusion as $T$
increases, the exact value of $C_{b}$ is not critical. However, the details
of the regime where more extensive phase diffusion takes place do depend on
the exact frequency dependence of the damping.

\begin{table}[tbp]
\caption{Total critical current ($I_{c0}$), sum of capacitances of two
junctions ($C$), inductance ($L$) and assymmetry of critical currents of the
two junctions of the DC-SQUID ($\protect\alpha $).}%
\begin{ruledtabular}
\begin{tabular}{ccccc}
DC-SQUID & $I_{c0}$ ($\mu$A) & $C$ (fF) & $L$ (pH) & $\alpha$  \\ \hline
A & 4.25 & 90 & 37 & 0.025 \\
B & 2.90 & 260 & 70 & 0.05
\end{tabular}
\end{ruledtabular}
\end{table}

The samples we used in this study were DC-SQUIDs fabricated using a
self-aligned Nb trilayer process \cite{Wei}. The inductances of the SQUIDs
were chosen small enough that $\beta _{L}=2\pi LI_{c0}/\Phi _{0}<<1$. So,
for most flux biases, $\Phi _{b}$, the potential of the SQUIDs could be
approximated well by that of a single junction with an effective critical
current $I_{c}(\Phi _{b})$ (Fig. 1c). Two samples with different parameters
were measured in our two laboratories. The results of these two measurements
are essentially identical. The parameters of the samples are listed in Table
1. We determined the total critical current of the SQUIDs by fitting the
switching rates in the temperature interval 0.4-1.0 K, where, for $\Phi
_{b}=0,$ the rates followed the Kramers rate (Eq. 1) well. We estimate the
total capacitance $(C)$ of two junctions of the SQUID from their area and
the specific capacitance of 45 fF/$\mu $m$^{2}$ and calculate the
inductances ($L$) by using a 3D inductance calculation program \cite{Khapaev}%
.

The sample is placed in a copper can filled with He exchange gas and located
on a temperature regulated stage of a dilution refrigerator. Extensive
electrical filtering is used to prevent interference and higher temperature
thermal noise from reaching the sample. The bias current of the SQUID is
ramped at constant rate. When a voltage appears across the device, the value
of bias current at which switching occurred is recorded. This is typically
repeated over 10$^{4}$ times at a given temperature and flux bias and a
histogram of the switching currents is compiled. In Fig. 3, typical
switching rates versus the ratio of barrier height to temperature from
sample $A$ are plotted. Here the barrier height is calculated from the full
two dimensional (2D) potential of the DC-SQUID for each value of bias
current. According to Eq. (1) the rates should fall onto one straight line
in this plot if the ln$(\omega _{p})$ dependence of rate on the bias current
is neglected. In the temperature range 0.4..1.8 K this is indeed the case.
At higher temperatures and larger value of $\Delta U/k_{B}T$, however, the
switching rates increasingly deviate from the simple exponential dependence
described by Eq. (1). Following our earlier argument we expect this to
happen when retrapping takes place. To verify this, we also plot the
modified rate calculated from simulations of Eq. (2) in the same figure. To
achieve a reasonable speed, a 1D approximation to the full 2D potential with
an effective $I_{c}(\Phi _{b})$ is used to obtain $P_{rt}$. The agreement
between measurement and the model is very good despite the simplicity of the
circuit model. 
\begin{figure}[t]
\centerline{\includegraphics[width=82mm]{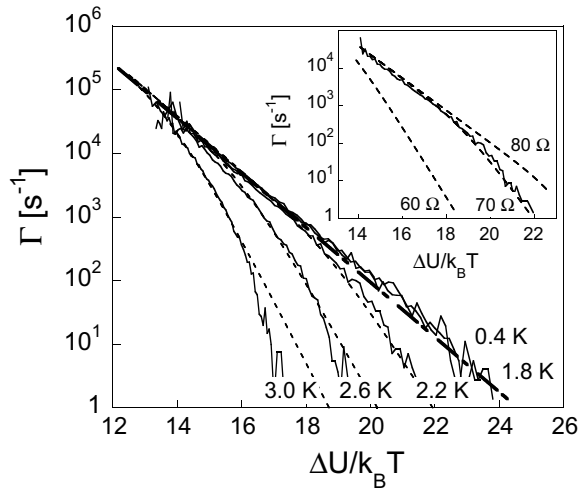}}
\caption{Switching rate of DC-SQUID $A$ as a function of barrier height to
temperature ratio (solid lines) for temperatures 0.4, 1.8, 2,2, 2.6, 3.0 K
and flux bias of DC-SQUID $\Phi _{b}=0.167$ $\Phi _{0}$. Dash-dot line shows
prediction of Eq. (1). Dashed lines represent prediction of Eq. (2) for
temperatures 1.8, 2.2, 2.6, 3.0 K. Note that predictions by eq. (1) and (2)
coincide for temperature 1.8 K and lower. Retrapping probability is
calculated for the model with parameters $R_{s}=70$ $\Omega $ and $C_{b}=20$
pF. Inset shows measured switching rate at temperature 2.2 K (solid line).
Dashed line show calculated rates for $R_{s}$ = 60, 70 and 80 $\Omega $.}
\end{figure}

Next we study the crossover to phase diffusion as a function of temperature
and $I_{c}(\Phi _{b}).$ For clarity, in this large data set, only the first
two moments of switching histograms, $\overline{I_{b}}$ and $\sigma _{I}^{2},
$ rather than the complete $\Gamma (I_{b})$ curves, are presented. Figure 4.
shows $\overline{I_{b}}$ and $\sigma _{I}$ for different values of $%
I_{c}(\Phi _{b})$ for sample $B$. One can notice three distinct regions in
the temperature dependence of the width $\sigma _{I}$. For flux bias $\Phi
_{b}\thickapprox 0$ and $T<80$ mK the escape is by quantum tunneling (QT)
giving a width that is nearly independent of $T$. For higher $T$ , in the
range of $0.1-2.0$ K, and for larger values of $I_{c}(\Phi _{b})$ $\ \sigma
_{I}$ agrees well with the predictions of TA (Eq. 1).The cross-over
temperature to QT decreases with $I_{c}(\Phi _{b})$, as expected, with the
very lowest $I_{c}$ data never completely reaching the QT regime.\ At still
higher temperatures, we observe a third regime where the width of the
distribution decreases as temperature is increased . Based on our earlier
discussion, this results from retrapping of the system after its initial
escape \cite{Kivioja}. Similarly to the second moment, $\overline{I_{b}}$
agrees well with theory in the QT and TA regimes. It is greater than
predicted by TA after the onset of retrapping since a greater tilt to the
potential is required to overcome the retrapping and initiate switching.

\begin{figure}[tbp]
\centerline{\includegraphics[width=85mm]{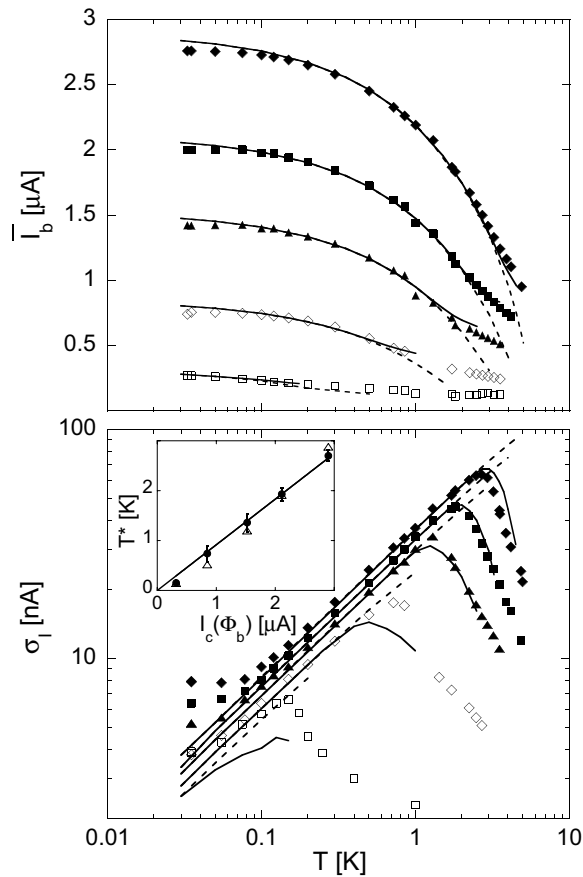}}
\caption{ Mean (top panel) and width (bottom panel) of switching
distribution as a function of temperature for sample $B$. Solid diamonds
correspond to flux bias $\Phi _{b}=0$ $\Phi _{0}$, solid rectangles to $\Phi
_{b}=0.239$ $\Phi _{0}$, solid triangles to $\Phi _{b}=0.324$ $\Phi _{0}$,
empty diamonds to $\Phi _{b}=0.409$ $\Phi _{0}$ and empty rectangles to $%
\Phi _{b}=0.474$ $\Phi _{0}$. Dashed lines show mean and width calculated
using rate given by Eq. (1). Solid lines represent results of calculation
based on Eq. (2). Inset shows temperature where the width of switching
distribution peaks as a function of effective critical current of the
DC-SQUID (circles) and results of calculation based on Eq. (2) (triangles).
Ramp rate of bias current is 88 $\protect\mu $A/s.}
\end{figure}

Figure 4 also shows calculations based on Eq. (2) (solid lines). As one can
see Eq. (2) predicts well the anomalous temperature dependence of the width
for $\Phi _{b}=0,$ 0.239 and 0.324 $\Phi _{0}$ but gives only a fair fit for 
$\Phi _{b}=0.409$ and 0.474 $\Phi _{0}$. For $\Phi _{b}\approx 0.5$ $\Phi
_{0}$, the 1D potential of a single junction with effective critical current 
$I_{c}(\Phi _{b})$ is a rather poor approximation to the actual 2D potential
of the DC-SQUID which can explain the discrepancy. Nevertheless, for all
values of effective critical currents, calculations based on Eq. (2) predict
reasonably accurately the temperature $T^{\ast }$at which the width of the
switching distribution is a maximum. The inset of Fig. 4. shows $T^{\ast }$
as a function of the effective critical current along with predictions of
Eq. (2). As this figure shows, $T^{\ast }(I_{c})$ spans more than one order
of magnitude for a given SQUID and scales linearly with the effective
critical current. Our calculations also show that the observed linear $%
T^{\ast }$ vs. $I_{c}(\Phi _{b})$ dependence follows only for sufficiently
small values of high frequncy damping, i.e. large $R_{s}$. For larger
damping, this dependence deviates from linear and approaches quadratic. The
simulations of dynamics of the phase particle (near $\Phi _{b}=0$) show also
that at the temperature $T^{\ast }$ the retrapping probability around the
mean of the switching distribution is relatively low ($P_{rt}(\overline{I_{b}%
})\thicksim 0.1$), so most escapes of the phase particle from local minima
lead to switching. In those cases when retrapping occurs, the number of
wells the phase particle moves before retrapping is, on average 2..3$.$ As
the temperature rises further and the average switching current decreases
advances of the phase particle before retrapping become shorter due to
smaller energy input from bias source but more frequent due to the increased
level of current fluctuations. This leads eventually to the appearance of a
stable phase diffusion state where the phase particle diffuses down the
tilted Josephson potential with uniform average speed giving rise to a
measurable phase diffusion voltage. We observe such a phase diffusion
voltage ($\thicksim $1 $\mu $V) for $\Phi _{b}=0.409$ and 0.474 $\Phi _{0}$
(sample $B$) at $T\thickapprox 4.2$ \ K.

In summary, we observe the crossover from Kramers escape to escape affected
by phase diffusion in relatively large Josephson junctions having $I_{c}$ of
the order of 1 $\mu $A. This is manifested by a peak in the width of the
switching distribution $\sigma _{I}$ vs T at a temperature $T^{\ast }(I_{c})$
which scales linearly with $I_{c}$. For $T>T^{\ast }(I_{c})$, $\sigma _{I}$
can decrease significantly below that expected for Kramers escape. Our data
are in good agreement with the results of Monte Carlo simulations based on a
simplified circuit model for the junction in which the higher damping at the
plasma frequency is represented by a series RC shunt.

The authors are grateful to D. V. Averin and K. K. Likharev for useful
discussions. Work is supported in part by NSF (DMR-0325551) and by AFOSR,
NSA and ARDA through DURINT grant (F49620-01-1-0439).

\end{document}